\documentclass[a4paper,11pt]{article}
\usepackage{pos}
\usepackage{wrapfig}
\usepackage{multicol}
\usepackage{slashed}
\usepackage{physics}
\usepackage{subcaption}

\title{Lattice QCD with an inhomogeneous magnetic field background}

\author[a]{B. B. Brandt}
\author[b]{F. Cuteri}
\author[a]{G. Endr\H{o}di}
\author[a]{G. Mark\'o}
\author*[a]{A. D. M. Valois}

\affiliation[a]{Universität Bielefeld,\\
  Universitätsstraße 25, 33615 Bielefeld, Germany}

\affiliation[b]{Institute for Theoretical Physics, Goethe University,\\
 Max-von-Laue-Straße 1, 60438 Frankfurt, Germany}

\emailAdd{dvalois@physik.uni-bielefeld.de}
\emailAdd{endrodi@physik.uni-bielefeld.de}
\emailAdd{brandt@physik.uni-bielefeld.de}
\emailAdd{gmarko@physik.uni-bielefeld.de}
\emailAdd{cuteri@itp.uni-frankfurt.de}

\newcommand{\ave}[1]{\left\langle\hspace{0.1cm}#1\hspace{0.1cm}\right\rangle}

\abstract{The magnetic fields generated in non-central heavy-ion collisions are among the strongest fields produced in the Universe, reaching magnitudes comparable to the scale of the strong interactions. Backed by model simulations, the resulting field is expected to be spatially modulated, deviating significantly from the commonly considered uniform profile. To improve our understanding of the physics of quarks and gluons under such extreme conditions, we use lattice QCD simulations with $2+1$ staggered fermion flavors with physical quark masses and an inhomogeneous magnetic background for a range of temperatures covering the QCD phase transition. We assume a $1/\cosh^2$ function to model the field profile and vary its strength to analyze the impact on the computed observables and on the transition. We calculate local chiral condensates, local Polyakov loops and estimate the size of lattice artifacts. We find that both observables show non-trivial spatial features due to the interplay between the sea and the valence effects.}
\FullConference{%
 The 38th International Symposium on Lattice Field Theory, LATTICE2021
  26th-30th July, 2021
  Zoom/Gather@Massachusetts Institute of Technology
}
\begin{document}
\maketitle
\section{Introduction}
Several physical systems are associated with the strongest magnetic fields that we know of in the Universe. Magnetars, for instance, are strongly magnetized neutron stars whose hot and dense cores can bear stable fields up to $10^{15}$ G ($\sqrt{eB}\sim1$ MeV) \cite{duncan1992formation}. Experiments of non-central heavy-ion collisions can produce transient fields of $10^{18}$ - $10^{19}$ G ($\sqrt{eB}\sim0.1$ - $0.5$ GeV) for RHIC and LHC energies, respectively \cite{skokov2009estimate}. Furthermore, cosmological models predict even higher fields during the electro-weak phase of the early Universe, when the primordial field could have reached magnitudes as high as $10^{20}$ G ($\sqrt{eB}\sim1.5$ GeV) \cite{vachaspati1991magnetic}. Since the strength of these fields is comparable to the energy scale of the strong interactions, the understanding of QCD in the presence of strong magnetic fields is crucial to answering questions about how quarks and gluons behave in high-energy collisions and the origin of galactic fields inherited from the early Universe.

In this work, we will focus on the implications of magnetic fields in the context of heavy-ion collisions. The case of strong uniform fields has been widely studied both numerically on the lattice (e.g. \cite{bali2012qcd,d2013lattice}) and analytically via QCD models (e.g. \cite{andersen2016phase}). However, in heavy-ion collision experiments, the resultant fields clearly deviate significantly from the uniform case. Since the magnetic field is non-uniform and changes rapidly with time (within $\sim1$ fm/c), it also generates a non-uniform time-dependent electric field which plays an important role in the dynamics of the byproducts of the collision and might have an impact on the phase transition. Event-by-event simulations of heavy-ion collisions predicted highly complex profiles for both the electric and magnetic field components \cite{voronyuk2011electromagnetic,deng2012event}. 
These facts bring about two main computational difficulties. 1) Real electric fields give rise to a sign problem, preventing direct simulations on the lattice. 2) The Minkowski time evolution of the fields is not attainable from Euclidean simulations. Bearing in mind these caveats, here we show how to improve our description of the complex heavy-ion collision scenario by implementing a non-uniform background magnetic field $B(x)$ in lattice QCD simulations. Our choice as $1/\cosh^2(x)$ for $B(x)$ is motivated by the profiles obtained from the aforementioned heavy-ion simulations, as well as due to the possibility of an analytical treatment of the free Dirac operator in this case~\cite{Dunne:2004nc,cao2018chiral}.
\par This contribution is organized as follows: in section \ref{sec:mag_field} we discuss some basic aspects of magnetic fields on the lattice, reviewing the flux quantization for the homogeneous case and introducing the $1/\cosh^2$ case. In section \ref{sec:results} we present our results for the local chiral condensate and local Polyakov loop. Finally, we summarize in section \ref{sec:conclusions}.
\section{Magnetic fields on the lattice}
\label{sec:mag_field}
\begin{figure}
    \centering
    \begin{subfigure}{0.49\textwidth}
    \includegraphics[width=\linewidth]{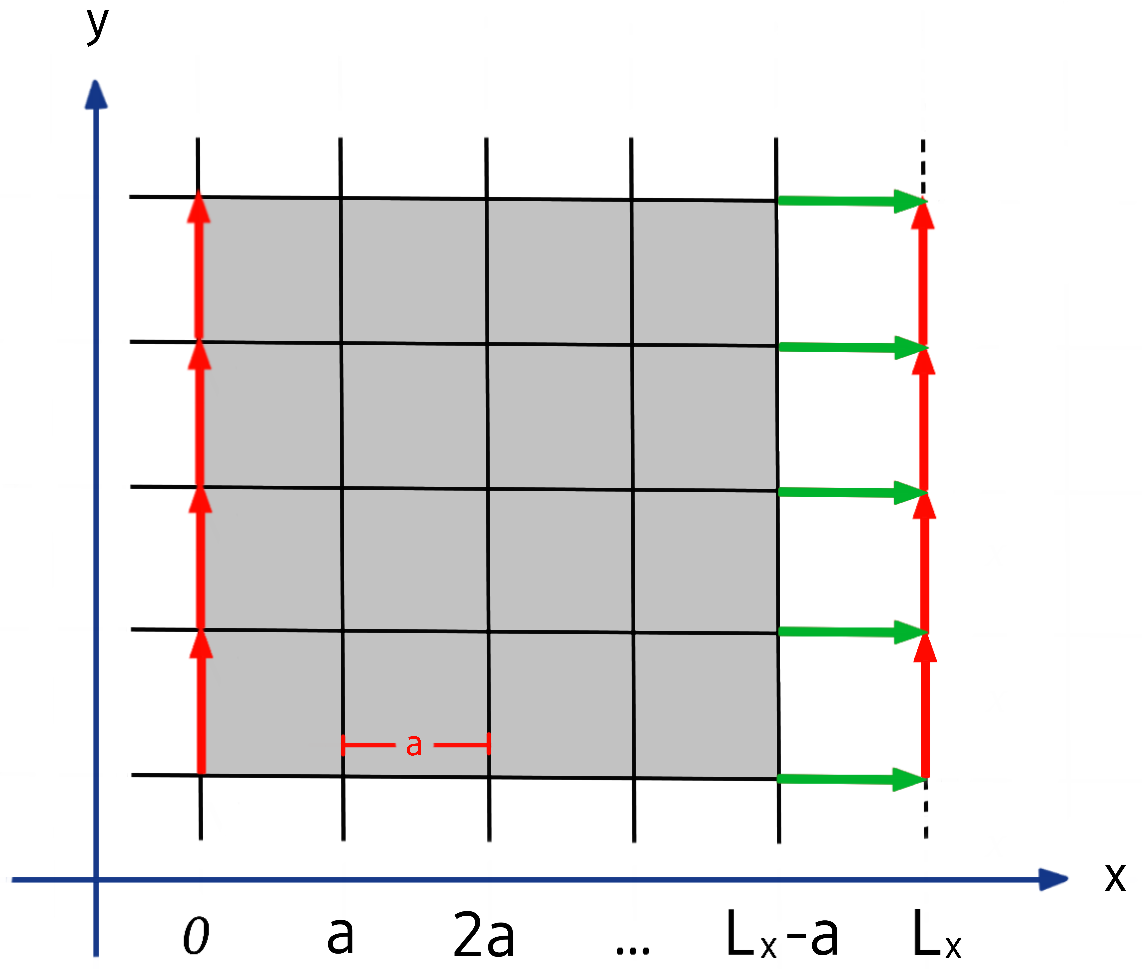}
    \end{subfigure}
    \begin{subfigure}{0.49\textwidth}
    \includegraphics[width=\linewidth]{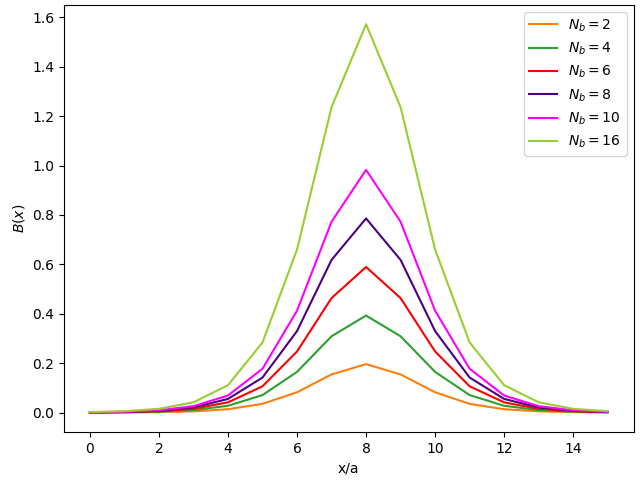}
    \end{subfigure}
    \caption{Left: $xy$-lattice plane with U(1) links satisfying periodic boundary conditions. Right: Profile of the $B$ field implemented on the lattice.}
    \label{fig:lattice + B-profile}
\end{figure}

In order to implement a magnetic field on the lattice, besides the non-Abelian SU(3) links corresponding to the gluon fields in QCD, we must also introduce Abelian links $u_{\mu}\in$ U(1) representing the magnetic field. For a homogeneous field pointing in the $z$ direction, a simple choice of links would be $u_y = e^{iaqBx}$ and $u_x=u_z=u_t=1$. However, it is convenient to have the U(1) links satisfy periodic boundary conditions. Therefore, we perform a gauge transformation on the $y$-links in the $(L_x,y)$-slice of the lattice, as depicted in the left plot of Figure \ref{fig:lattice + B-profile}. Since the transformation also affects the $x$-links in the $(L_x-a,y)$-slice, we must transform them in order to make them periodic. This series of transformations leads us to the following prescription for the links in the uniform case~\cite{bali2012qcd}
\begin{align}
u_{x}(x,y,z,t) &=
    \left\{
        \begin{array}{ll}
        e^{-iqBL_xy} & \mbox{if } x = L_x-a \nonumber\\
        1 & \mbox{if } x \neq L_x-a
        \end{array}
    \right. \\
u_{y}(x,y,z,t) &= e^{iaqBx} \\
u_z(x,y,z,t) &= 1 \nonumber\\
u_t(x,y,z,t) &= 1. \nonumber
\end{align}
The periodicity of the lattice imposes that the magnetic flux has to be quantized according to
\begin{equation}
qB = \frac{2\pi N_b}{L_xL_y},\hspace{1cm} N_b\in\mathbb{Z}.
\end{equation}
The same procedure can be applied for an inhomogeneous field of the form
\begin{equation}
\textbf{B} = \frac{B}{\cosh^2\qty(\frac{x-L_x/2}{\epsilon})}\hat{z}
\label{eq:inv_cosh_profile}
\end{equation}
where $\epsilon$ is the width of the $1/\cosh^2$ profile, centered in the middle of the lattice. The prescription for the links in this case is
\begin{align}
u_{x}(x,y,z,t) &=
    \left\{
        \begin{array}{ll}
        e^{-2iqB\epsilon y\tanh(\frac{L_x}{2\epsilon})} & \mbox{if } x = L_x-a \nonumber \\
        1 & \mbox{if } x \neq L_x-a
        \end{array}
    \right. \\
u_{y}(x,y,z,t) &= e^{iqB\epsilon a\qty[\tanh(\frac{x-L_x/2}{\epsilon}) + \tanh(\frac{L_x}{2\epsilon})]} \\
u_z(x,y,z,t) &= 1 \nonumber\\
u_t(x,y,z,t) &= 1. \nonumber
\end{align}
Similarly to the case of a uniform magnetic field, the flux of the inhomogeneous field is also quantized
\begin{equation}
qB = \frac{\pi N_b}{L_y\epsilon\tanh(L_x/2\epsilon)},\hspace{1cm} N_b\in\mathbb{Z}.
\label{eq:B-quantization-rule}
\end{equation}
The right plot in Figure \ref{fig:lattice + B-profile} shows the profile from Eq.~\eqref{eq:inv_cosh_profile} using the quantization rule from Eq.~\eqref{eq:B-quantization-rule} for different $N_b$ values.
\section{Results}\label{sec:results}
\begin{figure}[!h]
    \centering
    \begin{subfigure}{0.49\textwidth}
    \includegraphics[width=\linewidth]{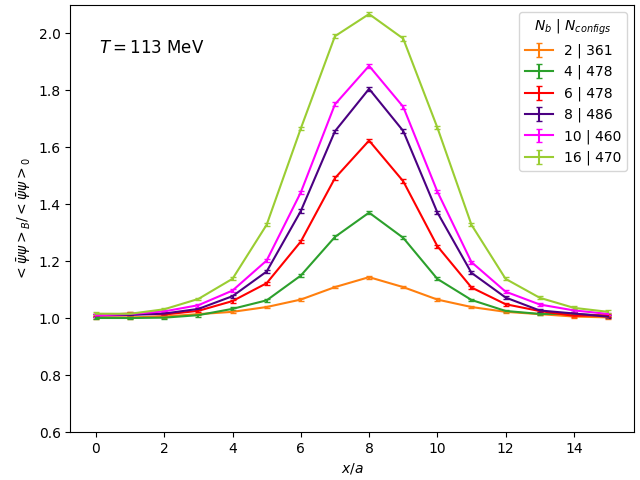}
    \end{subfigure}
    \begin{subfigure}{0.49\textwidth}
    \includegraphics[width=\linewidth]{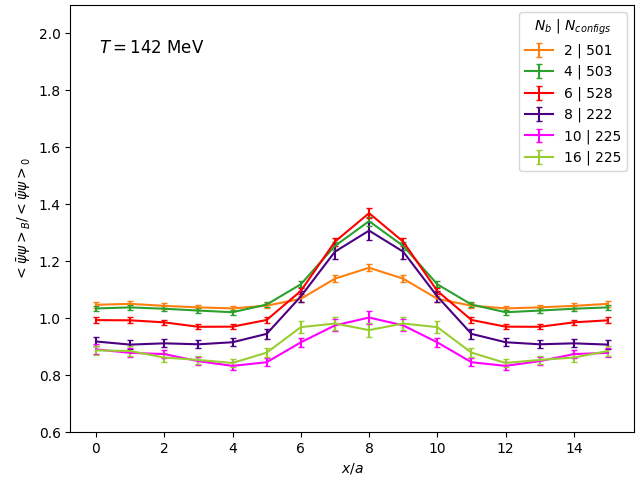}
    \end{subfigure}
    \begin{subfigure}{0.49\textwidth}
    \includegraphics[width=\linewidth]{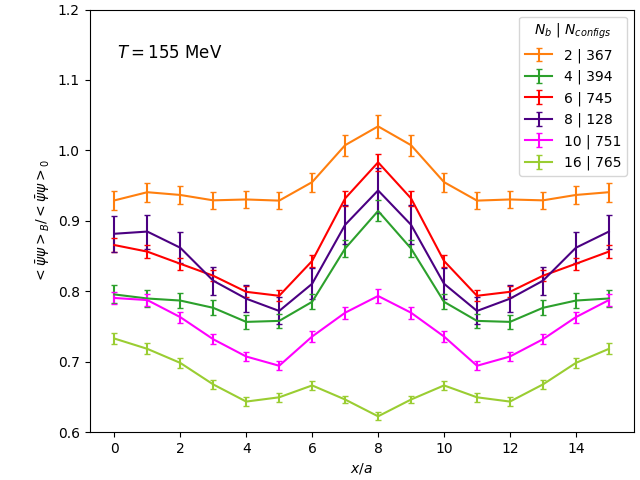}
    \end{subfigure}
    \begin{subfigure}{0.49\textwidth}
    \includegraphics[width=\linewidth]{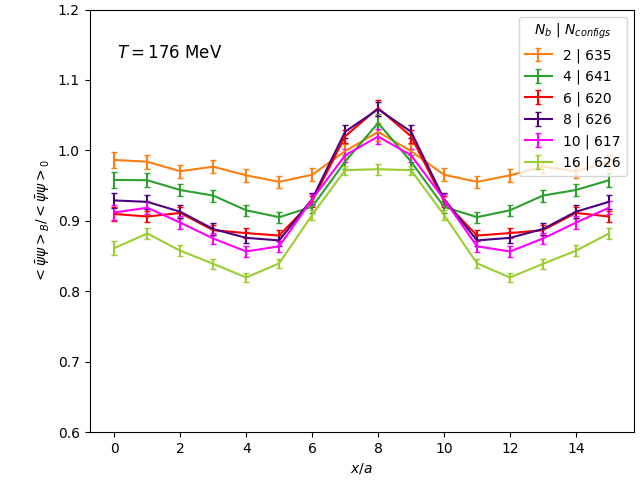}
    \end{subfigure}
    \caption{Average of the $u$- and $d$-condensates as a function of the $x$ coordinate (in lattice units) for different values of $N_b$ and temperature. The condensates are normalized by the zero-field condensate at each $T$. Beside each $N_b$, we indicate the number of sampled configurations.}
    \label{fig:local-condensates}
    \vspace{-0.5cm}
\end{figure}

Using $N_f = 2+1$ flavors of staggered fermions with physical quark masses, we generated gauge configurations on a $16^3\times6$ lattice for several values of the coupling $\beta$ and several values of the magnetic quantum number, namely $N_b = 0,2,4,6,8,10,16$, assuming a background field given by Eq.~\eqref{eq:inv_cosh_profile}. We fixed the width of the profile as $\epsilon/a=2$, which produced an appreciable inhomogeneity in the field on the lattice. In Table \ref{tab:parameters}, we show the set of parameters used for each $N_b$ as well as the temperature corresponding to each $\beta$ and the input quark masses in lattice units along the line of constant physics~\cite{Borsanyi:2010cj}. We chose couplings to have temperatures in the range from below $T_c$ to above $T_c$, where $T_c\sim155$ MeV is the crossover temperature. For each $B$ and $T$, we computed the local chiral condensate and the local Polyakov loop as
\begin{align}
\ave{\bar{\psi}\psi(x)}_B &= \frac{1}{Z}\int \mathcal{D}U\hspace{0.1cm} e^{-S_g}\det[\slashed{D}(x,B)+m]^{1/4}\Tr[\slashed{D}(x,B)+m]^{-1} \label{eq:chiral-condensate}\\
\ave{P(x)}_B &= \frac{1}{Z}\int \mathcal{D}U\hspace{0.1cm} e^{-S_g}\det[\slashed{D}(x,B)+m]^{1/4}\Re\Tr[\prod_{t}U_t(x)]. \label{eq:polyakov-loop}
\end{align}
\begin{table}
    \centering
    \vspace{-0.5cm}
    \begin{tabular}{l|l|l|l|l}
         $\beta$ & $T$ (MeV) & $a$ (fm) & $a\cdot m_{ud}$ & $a\cdot m_s$ \\
         \hline
         3.450	& 113 & 0.290 &	0.0057 & 0.163 \\
         3.525	& 142 & 0.232 &	0.0040 & 0.112 \\
         3.555 & 155 & 0.213 & 0.0035 &	0.098 \\
         3.600 & 176 & 0.187 &	0.0029 & 0.082 \\
    \end{tabular}
    \caption{Columns are, from left to right: inverse coupling, temperature, lattice spacing, up / down quark and strange quark masses (in lattice units).}
    \label{tab:parameters}
    \vspace{-0.6cm}
\end{table}
To compute the right-hand side of Eq.~\eqref{eq:chiral-condensate} we applied the method of noisy estimators, where at every configuration we measured the condensate using $80$ random vectors for each quark flavor. Figure \ref{fig:local-condensates} shows the results for the average of the light quark condensates as a function of the $x$ lattice coordinate. The condensates at finite $B$ were normalized by the zero-field condensate $\ave{\bar{\psi}\psi}_0$ at each $T$.
To understand these results, let us first summarize what we know about the impact of uniform fields on the condensate. The magnetic field has a direct coupling to the valence quarks through the $\Tr[\slashed{D}(x,B)+m]^{-1}$ factor in Eq.~\eqref{eq:chiral-condensate}. This coupling tends to enhance the condensate (so-called valence effect), and the increase depends on the local strength of the field. Besides that, the magnetic field also affects the gluonic environment, changing the gauge configurations by means of the fermion determinant $\det[\slashed{D}(x,B)+m]$ (sea effect), and acts to diminish the condensate~\cite{Bruckmann:2013oba}. For low temperatures, namely $T < T_c$, the valence effect is dominant and the condensate increases as a function of $B$, thus enhancing chiral symmetry breaking. This is the so-called magnetic catalysis. However, for $T \approx T_c$ the sea effect dominates and the condensate is decreased, thus favoring chiral symmetry. This is called inverse magnetic catalysis. A competition between the two produces the non-trivial behavior observed in Figure \ref{fig:local-condensates}. In the upper left plot ($T\sim113$ MeV), we see that the condensates have a peak in the middle of the lattice (where the field is at maximum), which grows with increasing $N_b$. When the temperature increases up to $T_c$, an increase in the field tends to decrease the condensate, since the sea effect becomes dominant. In the upper right plot ($T\sim142$ MeV) for strong $B$ we notice that the ratio $\ave{\bar{\psi}\psi}_B/\ave{\bar{\psi}\psi}_0 < 1$, which means that the magnetic field is starting to suppress the condensate. This behavior is observed until $T$ around the crossover temperature, namely in the lower left plot ($T\sim155$ MeV). However, for even higher temperatures we observe that the condensate starts to grow again for increasing $B$. This happens because the dominance of the sea effect over the valence effect is a phenomenon associated with the QCD transition \cite{Bruckmann:2013oba}. Furthermore, in the tails of the curves, we observe the formation of dips in the condensate for larger temperatures, the most prominent being around the transition temperature $T=155$ MeV. These dips are triggered by the different length scales on which local magnetic fields affect sea and valence contributions, as we will see in the results for the Polyakov loop. To separate the contributions originating from the two effects, we also measured the $B\neq0$ operator on $B=0$ configurations, corresponding to the valence effect, and the $B=0$ operator on $B\neq0$ configurations, corresponding to the sea effect, for three temperatures: $T < T_c$, $T \sim T_c$ and $T > T_c$. These measurements correspond respectively to the following expectation values
\begin{align}
\ave{\bar{\psi}\psi(x)}_{\mathrm{valence}} &= \frac{1}{Z}\int \mathcal{D}U\hspace{0.1cm} e^{-S_g}\det[\slashed{D}(x,0)+m]^{1/4}\Tr[\slashed{D}(x,B)+m]^{-1} \label{eq:valence-chiral-condensate} \\
\ave{\bar{\psi}\psi(x)}_{\mathrm{sea}} &= \frac{1}{Z}\int \mathcal{D}U\hspace{0.1cm} e^{-S_g}\det[\slashed{D}(x,B)+m]^{1/4}\Tr[\slashed{D}(x,0)+m]^{-1}. \label{eq:sea-chiral-condensate}
\end{align}

\begin{wrapfigure}{r}{0.5\textwidth}
    \centering
    \vspace{-0.5cm}
    \begin{subfigure}{\linewidth}
    \includegraphics[width=\linewidth]{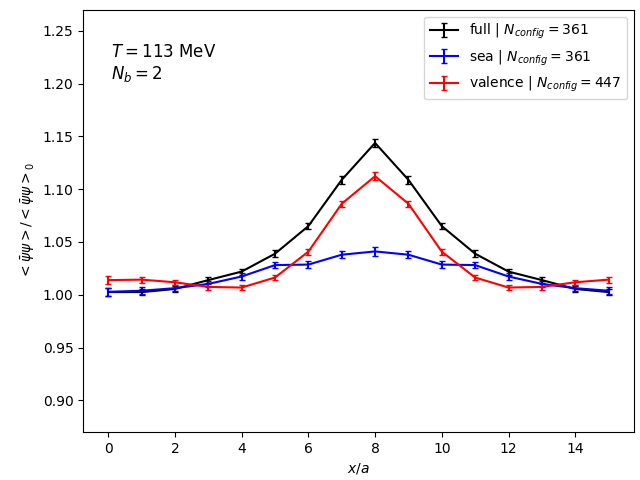}
    \end{subfigure}
    \begin{subfigure}{\linewidth}
    \includegraphics[width=\linewidth]{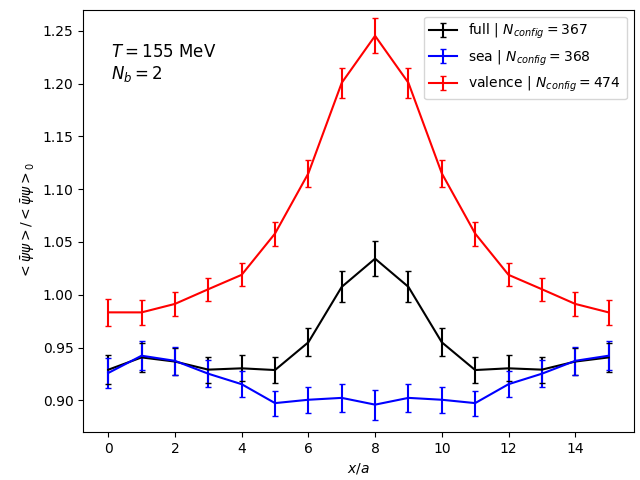}
    \end{subfigure}
    \begin{subfigure}{\linewidth}
    \includegraphics[width=\linewidth]{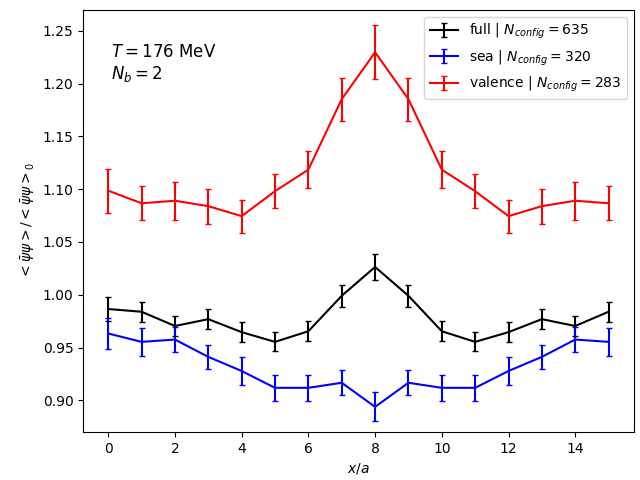}
    \end{subfigure}
    \caption{Contributions from the valence (red) and the sea (blue) effects to the $ud$-condensate at $N_b=2$ for three temperatures: one below $T_c$, one around $T_c$ and one above $T_c$. The curves are normalized by the zero-field condensate at each $T$}
    \label{fig:sea_valence_effects}
    \vspace{-0.2cm}
\end{wrapfigure}

\begin{figure}[!h]
    \centering
    \begin{subfigure}{0.49\textwidth}
    \includegraphics[width=\linewidth]{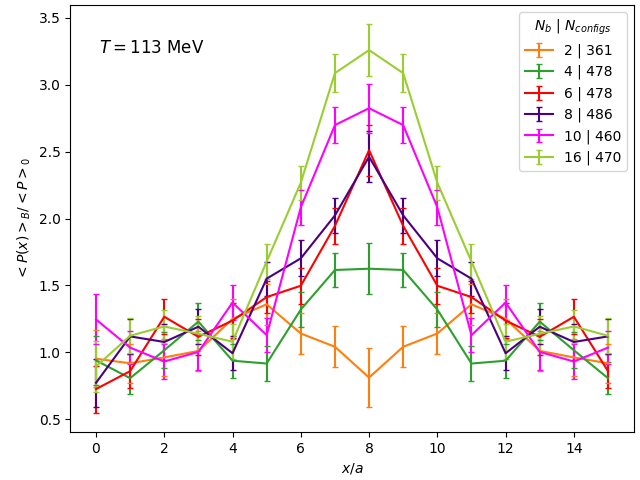}
    \end{subfigure}
    \begin{subfigure}{0.49\textwidth}
    \includegraphics[width=\linewidth]{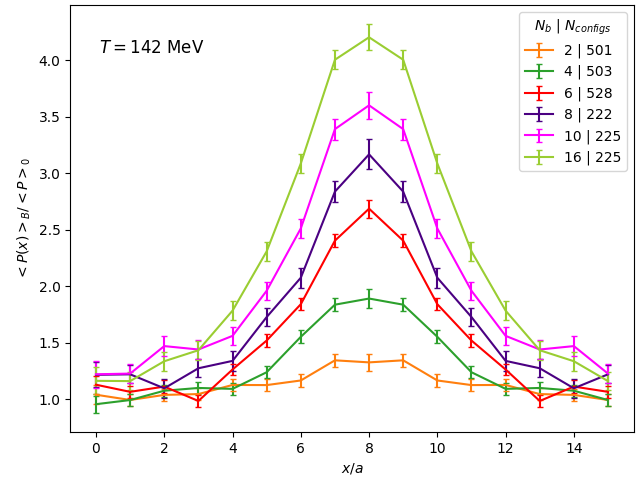}
    \end{subfigure}
    \begin{subfigure}{0.49\textwidth}
    \includegraphics[width=\linewidth]{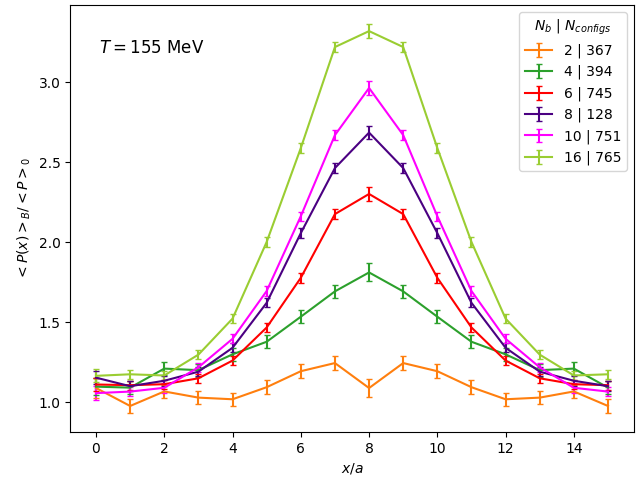}
    \end{subfigure}
    \begin{subfigure}{0.49\textwidth}
    \includegraphics[width=\linewidth]{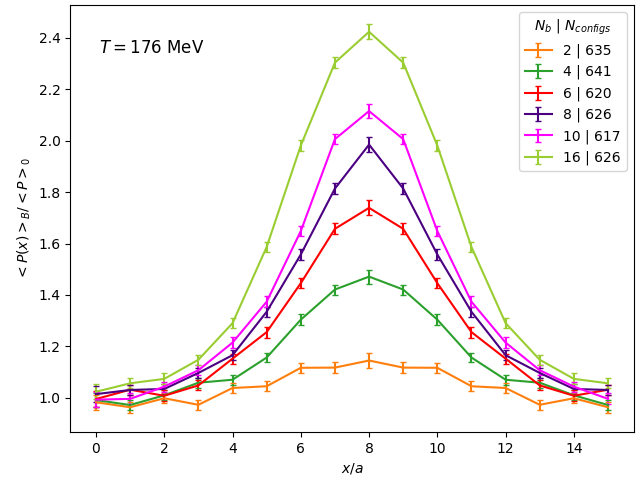}
    \end{subfigure}
    \caption{The Polyakov loop as a function of the $x$ coordinate (in lattice units) for different values of $N_b$ and the temperature. The Polyakov loops are normalized by the zero-field $P$ at each $T$. The temperature is indicated above each plot. Besides the $N_b$ values, we also show the number of sampled configurations.}
    \label{fig:polyakov-loops}
\end{figure}

The results for Eqs.~\eqref{eq:valence-chiral-condensate} and \eqref{eq:sea-chiral-condensate}, together with the full condensate, are shown in Figure \ref{fig:sea_valence_effects}. For low temperature (upper plot) the valence effect dominates and the condensate rises, assuming a similar shape to that of the pure valence profile (red curve). For a $T$ around the crossover temperature (middle plot), the situation is inverted, the sea effect becomes dominant and inverse magnetic catalysis takes place, causing the condensate to decrease in the region where $B$ is stronger. For a temperature above $T_c$ (lower plot) however, the valence contribution begins again to surpass the sea contribution and magnetic catalysis continues to increase the condensate. We emphasize that the tuning of the quark masses to their physical values is critical to the observation of inverse magnetic catalysis at the crossover temperature. Simulations with larger-than-physical quark masses, for instance, failed to reproduce the suppression of the condensate near that temperature and only magnetic catalysis was observed \cite{d2018qcd,endrHodi2019magnetic}. Indeed, this separation between valence and sea contribution makes sense and it can be interpreted in the following way. The right-hand side of Eq.~\eqref{eq:chiral-condensate}, for the full condensate, can be expanded in powers of $B$, where the first $B$-dependent term contains $B^2$, and we can see that the sea and the valence contributions appear within this expansion. Therefore, for low enough $B$, the valence and the sea terms are additive \cite{d2011chiral}. Besides the quark condensates, we also show the results for the Polyakov loop, in Figure \ref{fig:polyakov-loops}, since it is known to capture the most important changes in the gluonic fields due to the magnetic background~\cite{Bruckmann:2013oba}.
In this case, $P$ being a purely gluonic quantity, only the sea effect contributes. Since the weights $e^{-S_g}\det[\slashed{D}(x,B)+m]^{1/4}$ in the path integral depend on the magnetic field at all lattice points, this effect necessarily implies a smearing of the impact of $B$. Therefore we expect that the Polyakov loop is affected in a broader region of the lattice compared to the quark condensate, as we see in all the plots of Figure \ref{fig:polyakov-loops}. This difference in the length scales is what causes the dips on the tails of the condensates, as we mentioned above.
The presence of a background magnetic field tends to concentrate the lower Dirac eigenmodes around the lowest Landau level. This result is explained as a reduction in the dimensionality of degrees of freedom due to constraints of the motion of Dirac particles to the plane perpendicular to the magnetic field (see \cite{kharzeev2013strongly} for a detailed review), even in the presence of gluonic interactions~\cite{Bruckmann:2017pft}. In terms of the path integral, this consequence can be understood in the following way. Since the valence effect is related to the $\Tr[\slashed{D}(x,B)+m]^{-1}$ operator, when the eigenvalues are lowered, the trace of the inverse operator increases, thus increasing the condensate. On the other hand, the sea effect is related to a change in the gauge configurations by means of $\det[\slashed{D}(x,B)+m]$, for which a decrease in the eigenvalues tends to correlate with larger Polyakov loops, thus suppressing the condensate~\cite{Bruckmann:2013oba}.
\section{Summary and Outlook}\label{sec:conclusions}
In this proceedings article, we carried out a series of lattice QCD simulations, introducing an inhomogeneous magnetic field background using a $1/\cosh^2$ profile to model the fields appearing in heavy-ion collision simulations. We computed the local quark condensate and the local Polyakov loop for several sets of parameters. The quark masses were tuned to their physical value, and we argued that this was important to reproduce the correct behavior of the condensate in the transition region. We covered a range of temperatures from below to above the critical temperature to see the effect of an inhomogeneous field on the QCD phase transition. We showed that the quark condensate develops non-trivial features, especially towards the edges of the magnetic field peak, due to the interaction between the condensate and the Polyakov loop. We argued that these features are a manifestation of the interplay between the valence and the sea effects and computed the individual contributions from each, separately. In this study, we presented the first step towards more realistic modeling of the complex heavy-ion collision scenario and provided new insights on how important observables, like the quark condensate and the Polyakov loop, behave in such inhomogeneous backgrounds.
\\

\noindent
{\bf Acknowledgments}
This research was funded by the DFG (Emmy Noether Programme EN 1064/2-1 and the Collaborative Research Center CRC-TR 211 ``Strong-interaction
matter under extreme conditions'' -- project number
315477589 - TRR 211).
\bibliographystyle{utphys}
\bibliography{bibliography.bib}
\end{document}